
\input phyzzx

\input epsf

\pubnum{OU HEP 193}
\date{June, 1994}
\titlepage
\title{Quantum Vacuum and Anomalies}
\author{\rm Noriyuki Fumita\foot{e-mail: fumita@oskth.kek.jp}}
\address{\sl Department of Physics, Osaka University, Toyonaka, Osaka 560,
Japan}
\abstract{Chiral, conformal and ghost number anomalies are discussed from the
viewpoint of the quantum vacuum in Hamiltonian formalism.
After introducing the energy cut-off, we derive known anomalies in a new way.
We show that the physical origin of the anomalies is the zero point  
fluctuation
of bosonic or fermionic field.
We first point out that the chiral U(1) anomaly is understood as the creation
of the chirality at the bottom of the regularized Dirac sea in classical
electromagnetic field.
In the study of the (1+1) dimensional quantum vacuum of matter field coupled  
to
the gravity, we give a physically intuitive picture of the conformal anomaly.
The central charges are evaluated from the vacuum energy.
We clarify that the non-Hermitian regularization factor of the vacuum energy  
is
responsible for the ghost number anomaly. }

\endpage

\REF\Fuk{H. Fukuda and Y. Miyamoto, {\sl Prog. Theor. Phys.} {\bf 4} (1949)
347.}
\REF\Adl{S. L. Adler, {\sl Phys. Rev.} {\bf 177} (1969) 2426.}
\REF\Bel{J. S. Bell and R. Jackiw, {\sl Nuovo Cim.} {\bf 60A} (1969) 47.}

\REF\Fuj{K. Fujikawa, {\sl Phys. Rev. Lett.} {\bf 42} (1979) 1195; {\bf 44}
(1980) 1733; {\sl Phys. Rev.} {\bf D21} (1980) 2848.}
\REF\Alv{L. Alvarez-Gaume and P. Ginsparg, {\sl Nucl. Phys.} {\bf B243} (1984)
449.}
\REF\Nie{H. B. Nielsen and M. Ninomiya, {\sl Phys. Lett.} {\bf 130B} (1983)
389.}
\REF\Nel{P. Nelson and L. Alvarez-Gaume, {\sl Comm. Math. Phys} {\bf 99}  
(1985)
103.}
\REF\Niem{A. Niemi and G. W. Semenoff, {\sl Phys. Rev. Lett.} {\bf 55} (1985)
927.}
\REF\Son{H. Sonoda {\sl Nucl. Phys.} {\bf B266} (1986) 410.}
\REF\Pol{A. M. Polyakov, {\sl Phys. Lett.} {\bf 103B} (1981) 207, 211.}
\REF\Fuji{K. Fujikawa, {\sl Phys. Rev.} {\bf D25} (1982) 2584.}

\chapter{Introduction}
Since the Adler-Bell-Jackiw anomaly was discovered in the Feynman diagrammatic
calculation[\Fuk -\Bel], anomalies in quantum field theory have been discussed
from various viewpoints.
In path-integral formalism, the anomalies are identified with the Jacobian
factors under the classical symmetry transformations [\Fuj].
The path integral approach relates the chiral gauge anomalies to the
topological structures of the gauge theories[\Fuj][\Alv].
This approach gives a systematic understanding for all the anomalies.

In Hamiltonian formalism, the anomalies are discussed from the viewpoint of  
the
quantum vacuum.
The chiral U(1) anomaly is understood as the particle creation due to the
change of the Fermi surface of the massless Dirac sea in electromagnetic
field[\Nie].
This viewpoint is physically intuitive.
The topological analysis of the chiral anomalies is given in terms of the  
Berry
phase associated with the adiabatic change of the Dirac sea[\Nel -\Son].
However there are no systematic understanding of all the anomalies in the
Hamiltonian formalism.

In this paper, chiral, conformal and ghost number anomalies are discussed from
the viewpoint of the quantum vacuum in Hamiltonian formalism.
We show that the physical origin of the anomalies is the zero point  
fluctuation
of the bosonic or fermionic field.

In Sec.II, we consider the adiabatic change of the massive Dirac sea in
classical electromagnetic field.
We show that the chiral U(1) anomaly arises from the creation of the chirality
at the bottom of the regularized Dirac sea.
In this massive case, it is also shown that the chiral current conservation is
restored in the adiabatic process.
In Sec.III, we study the conformal anomaly in (1+1) dimensions.
The energy due to the zero point fluctuation in the quantum vacuum is
regularized keeping the reparametrization invariance.
We give a physically intuitive picture of the conformal anomaly.
The Liouville action and central charges are derived from the vacuum energy.
In Sec.IV, the ghost number current conservation
 is evaluated from the variation of the vacuum energy under the U(1)
transformation of (1+1) dimensional {\bf b}, {\bf c} system.
We point out that the non-Hermitian regularization factor of the vacuum energy
leads the ghost number anomaly.

\endpage

\chapter{Chiral U(1) anomaly}
In this section, we show how the chiral U(1) anomaly arises in the massive
Dirac sea in adiabatic process.

Let us start with the (1+1) dimensional Dirac fermion theory coupled to a
uniform electric field
$E=-{{\partial} \over {\partial t}} A^1(t)$ in the temporal gauge.
The Dirac Lagrangian is
$$
{\cal L}_{Dirac} = {\psi}^{\dag} (i {{\partial} \over {\partial t}} -
H[A^1(t)]) \psi , \eqn\DIRAC $$
$$
H[A^1(t)]=\sigma_1(-i {{\partial} \over {\partial x}} + eA^1(t)) + \sigma_3 m,
 \eqn\HAMILTON $$
where the signature of the metric is ($+$,$-$) with $t=x^0$, $x=x^1$,
and the gamma matrices are $\gamma^0=\sigma_3$, $\gamma^1=i\sigma_2$ and
$\gamma_5=\gamma^0 \gamma^1 = \sigma_1$.
Under the slowly varying potential $A^1(t)$, the Dirac sea changes
adiabatically.
We take adiabatic approximation within the order $eE/m^2$.

The Hamiltonian $H[A^1(t)]$ is diagonalized (within the order $eE/m^2$) as
$$ \eqalign{
H_{eff}&=UH[A^1]U^{\dag}-iU{\partial \over \partial t}U^{\dag} \cr
       &=\epsilon \sigma_3 + eE {m \over 2 \epsilon^2} \sigma_2,}  \eqn\DIAG
	   $$
where
$ \epsilon = \sqrt{m^2+\Pi^2} $, $U=(\sqrt{\epsilon + m}  + i\sqrt{\epsilon -
m} \sigma_2) / \sqrt{2 \epsilon} $,
$ \Pi = p + eA^1 $;
$p$ is the eigenvalue of the momentum $ -i \partial / \partial x $.
{}From \DIAG , we obtain
the eigenfunction of the energy $ \omega =\pm \epsilon ( \Pi )$,
$$ \eqalign{
U \psi_{+\epsilon (\Pi) } &= (1,ieEm/4\epsilon^3)^t e^{ipx} / \sqrt{L}, \cr
U \psi_{-\epsilon (\Pi) } &= (ieEm/4\epsilon^3,1)^t e^{ipx} / \sqrt{L}, }
\eqn\EIGEN
$$
where $L$ is the length of the system.
Since the `momentum' $\Pi(t)$ varies ($ \dot{\Pi} = -eE $), the eigenmodes  
flow
on the mass shell (Fig.1).
{}From exact calculation within the order $eE/m^2$,
we see the classical conservation law of the chiral current
$$
\partial_{\mu} j^{\mu}_{5 [\pm \epsilon ( \Pi )]} - 2im j_{5[\pm \epsilon (  
\Pi
)]} = 0, \eqn\CLASSICAL
$$
with
$j^{\mu}_{5[\pm \epsilon (\Pi )]}=\bar{\psi}_{\pm \epsilon (\Pi) } \gamma^\mu
\gamma_5 \psi_{\pm \epsilon (\Pi)} $,
$j_{5[\pm \epsilon (\Pi )]}=\bar{\psi}_{\pm \epsilon (\Pi) } \gamma_5  
\psi_{\pm
\epsilon (\Pi)} $.

\midinsert
$$\epsffile{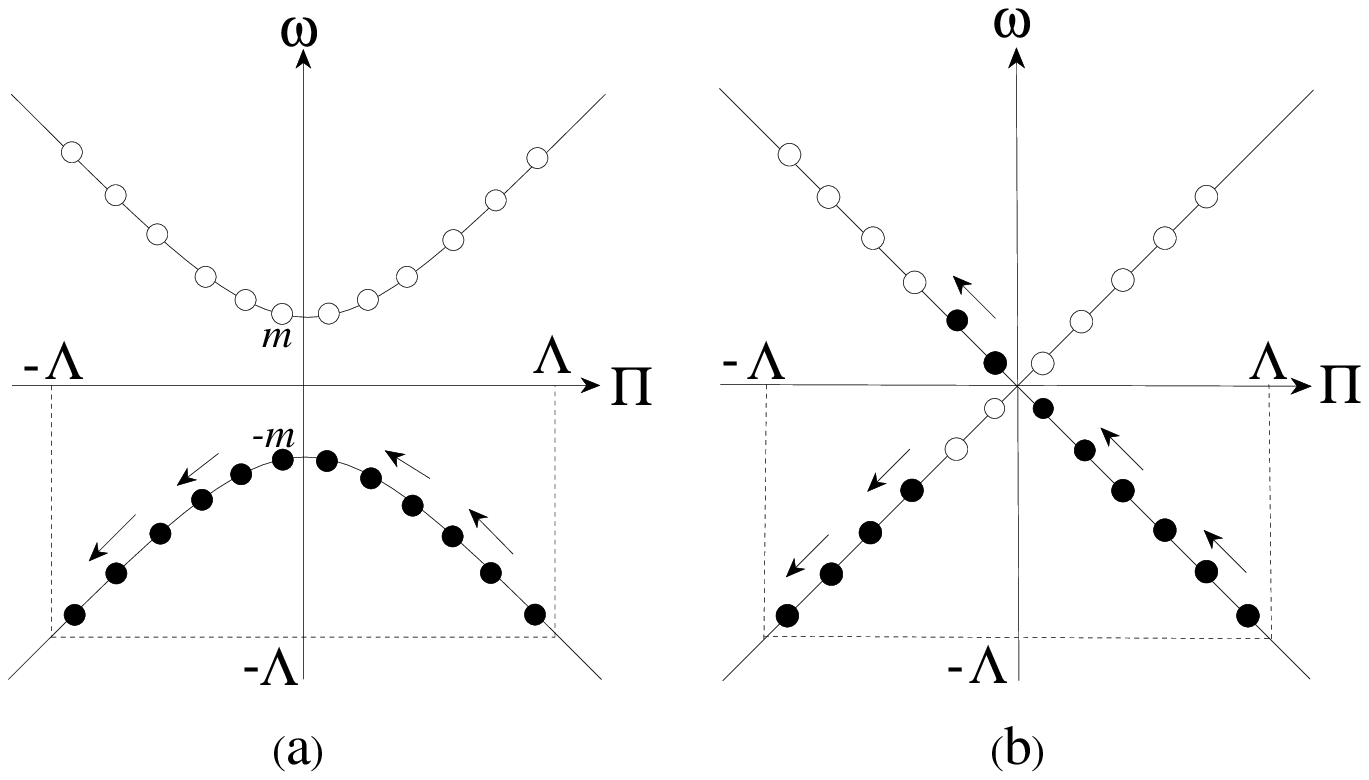}$$
\nobreak
\narrower
\singlespace
\noindent
Figure 1. Dispersion laws for (a)massive and (b)massless Dirac fermion in  
(1+1)
dimensions. The black and white points denote filled and unfilled modes
respectively. Each mode flows in the directions of arrows when the electric
field is turned on.
\medskip
\endinsert

The chiral current in the Dirac sea is the sum of all contribution from the
negative-energy modes.
We apparently obtain from \CLASSICAL\ the conservation law of chiral current  
in
the Dirac sea,
$$
\partial_\mu \Bigl{(} L \int {{d \Pi} \over {2 \pi}} j^{\mu}_{5 [- \epsilon
(\Pi) ]} \Bigr{)} - 2im \Bigl{(} L \int {{d \Pi} \over {2 \pi}} j_{5 [-
\epsilon (\Pi) ]} \Bigr{)} = 0. \eqn\SUM
$$
In this naive conservation law, however, each term in the left hand side is
ill-defined.
We regularize them by introducing the cut-off $\Lambda \gg m$;
$$ <j_5^{\mu}> = L \int {{d\Pi} \over {2 \pi}} j^{\mu}_{5[-\epsilon (\Pi)]}
\exp \big{(}-{{\Pi}^2 \over {\Lambda}^2}\big{)}. \eqn\REG $$
$<j_5>$ is defined similarly.
Note that
the regularized Dirac sea keeps the gauge invariance
as far as taking the cut-off on $\Pi$.

Now let us see the bottom of the Dirac sea (energy $\omega \sim - \Lambda $)  
in
Fig.1.
The negative energy electrons with chirality $(\Pi / \omega) \sim (-1)$ are
created
at the bottom $(\Pi, \omega) \sim (\Lambda ,-\Lambda)$.
The negative-energy electrons with chirality $(\Pi / \omega) \sim (+1)$ are
annihilated
at the bottom $(\Pi, \omega) \sim (-\Lambda ,-\Lambda)$.
{}From the equation of motion $ \dot{\Pi} = -eE $, the rate of the creation of
the chirality from the bottom is evaluated as $-2(eE/2\pi)L$ par unit time.
According to this anomalous effect, the current conservation \SUM\ is modified
to
$$
\partial_\mu <j_5^\mu> - 2im <j_5> = - {eE \over \pi}. \eqn\CHIRAL
$$
This is the anomalous chiral U(1) identity as is well known.

In the massless case ($m=0$), the Fermi surface changes and the electrons with
chirality (-1) and positrons with chirality (-1) are created in the electric
field(Fig.1-b). This creation breaks the chiral current conservation[\Nie].

In the massive case ($m \not= 0$), the Fermi surface does not change, thus the
chirality preserves in the adiabatic process(Fig.1-a).
This is independently verified from the Feynman diagrammatic calculation of  
the
vacuum expectation value $2im<j_5> \sim eE/\pi$ under the uniform and constant
electric field. The chiral current conservation is restored as \break
$\partial_\mu <j_5^\mu> = 2im<j_5> - eE/\pi \sim 0$ in the adiabatic process.

In (1+3) dimensions, we first consider the Dirac fermion in a uniform magnetic
field $B(>0)$ along the third direction.
We take the gamma matrices as $\gamma^0 = I \otimes \sigma_3$, $\gamma^k = i
\sigma_k \otimes \sigma_2$ and $\gamma_5 = i\gamma^0 \gamma^1 \gamma^2  
\gamma^3
= I \otimes \sigma_1$.
The energy levels are $ \omega = \pm \epsilon(\Pi,n) = \pm \sqrt{ m^2 + 2neB +
\Pi^2 } $ ($n=0,1,2,\cdots $) with $ \Pi = p^3 +e A^3 $.
For the non-zero mode ($ n \not= 0$), spin-up state and spin-down state
degenerate and the sum of their chirality becomes zero.
For the zero mode ($ n=0$), only spin-down mode exists.
Thus only negative-energy zero modes contribute the chirality of the Dirac  
sea.

Next a uniform electric field $E = -(\partial / \partial t) A^3$ is turned on
parallel to $B$, then $\Pi(t) = p^3 + eA^3(t)$ varies and the eigenmodes flow
on the shell $\omega (\Pi )=\pm \sqrt{m^2 + 2neB + \Pi^2}$.
Each mode satisfies the classical conservation law of chiral current.
We introduce the energy cut-off for the regularization. The chirality is
created from the bottom of the Dirac sea.
By using the chirality of the zero mode $(-\Pi / \omega)$ and the density of
the zero-mode states
 $(eB/2\pi)(L/2\pi)$ per length $L$, the rate of creation of the chirality is
evaluated as $e^2 EB/2\pi^2 $ per unit time and volume.
This is the chiral U(1) anomaly.
In this massive case, the Fermi surface does not change and the chirality is
preserved in the adiabatic process. This is independently verified from the
Feynman diagrammatic calculation of the vacuum expectation value $2im<j_5>  
\sim
-e^2 EB / 2 \pi^2 $, which is similar to the (1+1) dimensional case discussed
above.

\endpage

\chapter{Conformal anomaly in (1+1) dimensions}
In this section, we study the conformal anomaly in (1+1) dimensions from the
viewpoint of  the vacuum energy in the background metric.

We start from the scalar field $X(t,x)$ with the Lagrangian
$$
{\cal L} = {1 \over 2} \sqrt{-g(t,x)} g^{\mu \nu} (t,x) \partial_{\mu} X(t,x)
\partial_{\nu} X(t,x), \eqn\POLI
$$
where the signature of the metric is ($+$,$-$) with $t=x^0$, $x=x^1$.
Under the conformal transformation $g_{\mu \nu} \rightarrow e^{\phi (t,x)}
g_{\mu \nu} $, the Lagrangian \POLI\ does not change. The quantum vacuum
changes however.
To see this, we consider the vacuum energy due to the zero point fluctuation  
of
the scalar field.
We regularize the vacuum energy in flat space-time metric $\eta_{\mu \nu}$ as
$$
E[0] = L \int {dk \over 2 \pi} {1 \over 2} |k| \exp \big{(} -{k^2 \over
\Lambda^2} \big{)}, \eqn\ENERGY
$$
where $L$ is the length of the system.

The vacuum energy in the background metric $ e^{\phi} \eta_{\mu \nu} $ is
determined from the reparametri-zation invariance in the following way  
(Fig.2).
Under the reparametrization
$$ x^{\mu} \rightarrow {x^\prime}^{\mu} = e^{-\phi  / 2} x^{\mu},  \eqn\TRANSX
$$
the momentum is transformed as
$ k^{\mu} \rightarrow {k^{\prime}}^{\mu} = e^{\phi /2} k^{\mu}$, where
$\phi$ is a constant.
By noting $L \rightarrow L^{\prime} = e^{ - \phi /2} L $ and $ \Lambda
\rightarrow
{\Lambda}^{\prime} =e^{ \phi /2} \Lambda $, we obtain the transformation of  
the
vacuum energy
$$
E[0] \rightarrow {E[0]}^{\prime} = e^{-{\phi \over 2}} L \int {dk \over 2\pi}
{1 \over 2} |k| \exp \big{(} - {k^2 \over {e^{\phi} {\Lambda}^2}} \big{)}.
\eqn\ENERGYD
$$
Here we rescale the length of this system as
$$
 L \rightarrow e^{\phi  / 2} L,  \eqn\TRANSL
$$
then the vacuum energy \ENERGYD\ is transformed as
$$
E[0]^{\prime} \rightarrow E[\phi ] = L \int {dk \over 2\pi} {1 \over 2} |k|
\exp \big{(} - {k^2 \over {e^\phi \Lambda^2}} \big{)}.  \eqn\ENERGYP $$
Under the composite transformation of \TRANSX\ and \TRANSL ,
the metric tensor and domain of the parameter $x$ are transformed as
$$ \eta_{\mu \nu}   (0 \leq x \leq L)  \rightarrow  e^{\phi} \eta_{\mu \nu}(0
\leq x \leq e^{-\phi /2} L)  \rightarrow  e^{\phi} \eta_{\mu \nu}(0 \leq x  
\leq
L)
\eqn\METRICTR $$
which is equivalent to the conformal transformation $ \eta_{\mu \nu}
\rightarrow e^{\phi} \eta_{\mu \nu} $($0 \leq x \leq L$).
Thus the $E[\phi ]$ in \ENERGYP\ is the vacuum energy in the background metric
$ e^{\phi} \eta_{\mu \nu}$.

\midinsert
$$\epsffile{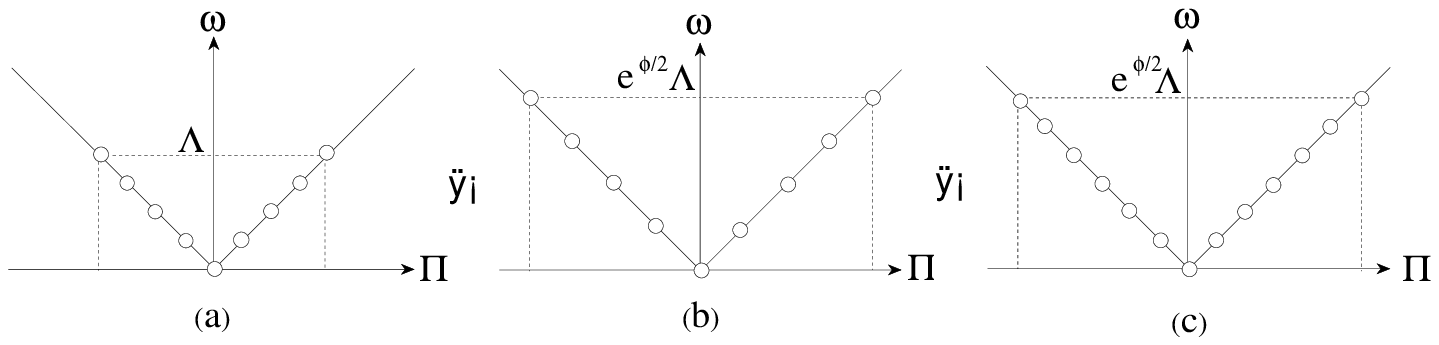}$$
\nobreak
\narrower
\singlespace
\noindent
Figure 2. Dispersion law for the scalar field in (1+1) dimensions. Under the
reparametrization $x^{\mu} \rightarrow {x^{\prime}}^{\mu} =e^{-\phi /2}  
x^{\mu}
$ and rescaling the length of this system $L \rightarrow e^{\phi / 2} L$, the
dispersion law is transformed as (a) $\rightarrow$ (b) and (b) $\rightarrow $
(c) respectively. In (a) $\rightarrow $ (c), each mode does not change, but  
the
cut-off level changes.
\medskip
\endinsert

In the background metric $e^{ \phi (x)} \eta_{\mu \nu}$ which depends on the
coordinate $x$ and is independent of the time $t$, the vacuum energy \ENERGYP\
is generalized as
$$
E[\phi (x)] = L \int {dk \over 2\pi} {1 \over 2}|k|\big{\langle} f_k , \exp
\big{(} -{\triangle \over \Lambda^2 } \big{)}~ f_k  \big{\rangle}_{scalar,}
\eqn\ENERGYPX $$
where
$f_k $ is the positive-energy wave function of momentum $k$,
$$
f_k  = e^{ikx - i|k|t}/\sqrt{2|k|L}.
\eqn\SCALWAVE $$
$\triangle$ is the one dimensional Laplacian $ \triangle = -|g^{11} (x) |
\nabla_1 \nabla_1 $ which is
$$ \triangle^{(j)} = - {1 \over e^{\phi}} \big{(} \partial_1 - {j+1 \over 2}
\partial_1 \phi \big{)} \big{(} \partial_1 - {j \over 2} \partial_1 \phi
\big{)}
\eqn\LAPJ
$$
for the field of conformal weight $j$, and the inner product  $\langle ~,~
\rangle_{scalar} $ is given by
$$
\big{\langle} f_l , e^{-\triangle / \Lambda^2} f_k \big{\rangle}_{scalar} =
\int dx [i f_l^{\ast} \exp \big{(} -{\triangle^{(1)} \over \Lambda^2} \big{)}
{\partial \over \partial t}f_k - i\big{(}{\partial \over \partial t}  
f_l^{\ast}
\big{)} \exp \big{(} - {\triangle^{(0)} \over \Lambda^2} \big{)} f_k].
\eqn\SCALARPRODR $$
Note that the regularization factor $ \exp (-\triangle / \Lambda^2) $ is
reparametrization invariant in one dimension.

{}From the momentum integration of \ENERGYPX, we obtain the vacuum energy
$$ \eqalign{
E[\phi (x)]
&= \int dx \int {dk \over 2\pi}{1 \over 2}|k| e^{-ikx} { \exp \big{(}
-{\triangle^{(1)} \over \Lambda^2} \big{)} +
 \exp \big{(} -{\triangle^{(0)} \over \Lambda^2} \big{)} \over 2} e^{ikx} \cr
&= \int dx \big{[} {\Lambda^2 \over 4\pi} e^\phi - {1 \over 96 \pi}(\partial_1
\phi)^2 - {1 \over 48\pi} {\partial_1}^2 \phi + O ({1 \over  
\Lambda^2})\big{]}.
}\eqn\ENERGYS $$
The third term ${\partial_1}^2 \phi $ is the total derivative term and gives  
no
contribution to the vacuum energy under the periodic boundary condition.
$O(1/\Lambda^2)$ vanishes by taking $\Lambda \rightarrow \infty $.

By noting the relation between the path-integral formalism and the Hamiltonian
formalism
$$
<0|\exp (-i \int dt H)|0> = \int {\cal D} X \exp (i \int d^n x {\cal L}),
\eqn\Relation $$
the partition function $Z_X [\phi (x)] $ is written as
$$
Z_X [\phi (x)] = \exp \big{( }  -i \int dt E[\phi (x)] \big{)}.  \eqn\PARTI
$$
Thus, under the conformal transformation $\eta_{\mu \nu} \rightarrow e^{\phi
(x)} \eta_{\mu \nu} $, the partition function is transformed as
$$
Z[0] \rightarrow Z[\phi (x) ] = Z[0] \exp (iS_L),  \eqn\TRANSZ  $$
$$
S_L
= \int dt \int dx \big{[} {1 \over 96 \pi} ( \partial_1 \phi)^2 - {\Lambda^2
\over 4\pi} (e^\phi - 1) \big{]}.  \eqn\LIOUV
$$
This $S_L$ is the Liouville action.
Equations \TRANSZ\ and \LIOUV\ justify that the quantum vacuum changes under
the conformal transformation.
The Liouville action denotes the difference of the vacuum energy: $S_L = -  
\int
dt (E[\phi (x)] - E[0])$.

{}From the relativistic generalization $-(\partial_1 \phi)^2 \rightarrow
(\partial_0 \phi)^2 - (\partial_1 \phi)^2 $ and Wick rotation $t \rightarrow
-i\tau$, $\Lambda \rightarrow iM$, we obtain the Euclid version of the
Liouville action[\Pol][\Fuji].
Here we note that the total derivative term ${\partial_1}^2 \phi$ in \ENERGYS\
gives no contribution to the conformal anomaly.
After the relativistic generalization $  -{\partial_1}^2 \phi \rightarrow
({\partial_0}^2 - {\partial_1}^2) \phi= -\sqrt{-g}R$ and Wick rotation, $
\partial_1^2 \phi $ term becomes the Einstein term $\int d^2 x \sqrt{g}R$.
The Einstein term gives the Euler number and does not change under the
conformal transformation.

For Majorana fermions, vacuum has negative energy and the vacuum energy  
formula
\ENERGYPX\ becomes
$$
E[\phi (x)] = -L \int {dk \over 2\pi} {1 \over 2}|k|\big{\langle} f_k , \exp
\big{(} -{\triangle^{(1/2)} \over \Lambda^2 } \big{)}~ f_k
\big{\rangle}_{spinor,}
\eqn\ENERGYPXM $$
where $f_k$ is the positive energy wave function of momentum $k$,
$$ f_k = e^{ikx - i|k|t}/\sqrt{L} \eqn\SPINWAVE $$
which is right (left) handed wave function for $k>0$ ($k<0$).
The conformal weight of the Majorana fermion is $1/2$, thus the one  
dimensional
Laplacian \LAPJ\ becomes $\triangle^{(1/2)}$. The inner product $\langle
{}~,~\rangle_{spinor}$ is given by
$$ \big{\langle} f_l,e^{-\triangle^{(1/2)}/\Lambda^2} f_k
\big{\rangle}_{spinor} = \int dx f_l^{\ast} \exp \big{(} -{\triangle^{(1/2)}
\over \Lambda^2 } \big{)} f_k . \eqn\SPINPROD $$
{}From the momentum integration of \ENERGYPXM , we obtain the vacuum energy
$$ \eqalign{
E[\phi (x)]
&=-\int dx \int {dk \over 2\pi}{1 \over 2}|k| e^{-ikx}  \exp \big{(}
-{\triangle^{(1/2)} \over \Lambda^2}  \big{)} e^{ikx} \cr
&= \int dx \big{[} -{\Lambda^2 \over 4\pi} e^\phi - {1/2 \over 96
\pi}(\partial_1 \phi)^2 + {1 \over 48\pi} {\partial_1}^2 \phi + O ({1 \over
\Lambda^2})\big{]}.
}\eqn\ENERGYSP $$
The Liouville action is derived from the first and second term.
In particular from the second term we obtain the central charge $1/2$.

\endpage

\chapter{Ghost number anomaly in (1+1) dimensions}
In this section, we study the vacuum energy of the {\bf b}, {\bf c} system and
the ghost number anomaly.

{}From the canonical quantization, the vacuum energy of the {\bf b}, {\bf c}
system becomes $E_{{\bf b}{\bf c}} =-\epsilon L \int (dk/2\pi)|k|$ with
$\epsilon = -1 $ ($+1$) for the bosonic (fermionic) ghost.
The bases of the wave function of the {\bf b}, {\bf c} system are taken as
$$ \eqalign{
f_k^{({\bf b})}  &= e^{ikx - i|k|t}/L^{\lambda},   \cr   f_k^{({\bf c})}  &=
e^{ikx - i|k|t}/L^{1-\lambda}, }\eqn\BCWAVE $$
where $\lambda$ (or $1-\lambda$) is the conformal weight of {\bf b} field (or
{\bf c} field).
The inner product  $\langle ~,~\rangle_{ghost}$ is given by
$$ \big{\langle} f^{({\bf b})}_l,e^{-\triangle^{(1-\lambda)}/\Lambda^2}
f^{({\bf c})}_k \big{\rangle}_{ghost} = \int dx {f^{({\bf b})}_l}^{\ast} \exp
\big{(} -{\triangle^{(1-\lambda)} \over \Lambda^2 } \big{)} f^{({\bf c})}_k .
\eqn\GHOSTPROD$$
Then, the vacuum energy is regularized as
$$
E_{{\bf b}{\bf c}} = -\epsilon L \int {dk \over 2\pi}|k| \big{\langle}
f_k^{({\bf b})}, \exp\big{(} - {\triangle^{(1-\lambda)} \over \Lambda^2}
\big{)} ~f_k^{({\bf c})}  \big{\rangle}_{ghost}
= \int dx {\cal E}_{\epsilon ,1-\lambda,}
\eqn\ENERGYC $$
$$
{\cal E}_{\epsilon ,1-\lambda}=  {-\epsilon \Lambda^2 \over 2\pi} e^{\phi} -
{-2\epsilon [6\lambda (\lambda-1) +1] \over 96 \pi} (\partial_1 \phi )^2
-\epsilon {-2+3(1-\lambda) \over 12\pi } {\partial_1}^2 \phi + O({1 \over
\Lambda^2}) .
\eqn\ENERGYDC $$
The regularization factor $\exp(-\triangle^{(1-\lambda)}/ \Lambda^2)$ is not
Hermitian for $\lambda \not= 1/2$.
{}From the definition of Laplacian $\triangle^{(j)}$ \LAPJ, it is shown that
$\triangle^{(1-\lambda )} $ and $\triangle^{(\lambda )}$ are Hermitian
conjugate.
By using this, the vacuum energy \ENERGYC\ is rewritten as
$$
E_{{\bf b}{\bf c}} = -\epsilon L \int {dk \over 2\pi}|k| \big{\langle}
\exp\big{(} - {\triangle^{(\lambda)} \over \Lambda^2} \big{)} ~f_k^{({\bf b})}
,f_k^{({\bf c})} \big{\rangle}_{ghost}
= \int dx {\cal E}_{\epsilon ,\lambda,}
\eqn\ENERGYB $$
$$
{\cal E}_{\epsilon ,\lambda}=  {-\epsilon \Lambda^2 \over 2\pi} e^{\phi} -
{-2\epsilon [6\lambda (\lambda-1) +1] \over 96 \pi} (\partial_1 \phi )^2
-\epsilon {-2+3\lambda \over 12\pi } {\partial_1}^2 \phi + O({1 \over
\Lambda^2}) .
\eqn\ENERGYDB $$
The difference between ${\cal E}_{\epsilon, 1-\lambda}$ and ${\cal
E}_{\epsilon, \lambda}$ is the total derivative term and this gives no
contribution to the vacuum energy $E_{{\bf b}{\bf c}}$.
{}From the second term of \ENERGYDC\ or \ENERGYDB, we obtain the familiar
central charge of the {\bf b}, {\bf c} system: $C_{ghost}= -2\epsilon[6\lambda
(\lambda -1) +1]$.

Next we consider the U(1) transformation of the {\bf b}, {\bf c} system
$$
\eqalign{
 {\bf b}(t,x) \rightarrow  e^{- \alpha (t,x)}{\bf b}(t,x) , \cr
 {\bf c}(t,x) \rightarrow  e^{+ \alpha (t,x)}{\bf c}(t,x) . }
\eqn\UBC $$
By noting the transformation of the wave function
$$ \eqalign{
f_k^{({\bf b})}  \rightarrow e^{-\alpha (t,x)}f_k^{({\bf b})} ,  \cr
f_k^{({\bf c})}  \rightarrow e^{+\alpha (t,x)}f_k^{({\bf c})} ,    }\eqn\UBCWF
$$
we obtain from
\ENERGYC\ and \ENERGYB\ the variation of the vacuum energy
$$
\delta E_{{\bf b}{\bf c}} = \int dx  \alpha (t,x) [ {\cal E}_{\epsilon
,\lambda} - {\cal E}_{\epsilon ,1-\lambda} ] \eqn\UBCE $$
for the infinitesimal parameter $\alpha (t,x)$.
{}From the relation \Relation, the infinitesimal transformation of the
effective action of the {\bf b}, {\bf c} system is evaluated as
$$
\delta \Gamma_{{\bf b}{\bf c}} = - \int dt \int dx \alpha (t,x) [ {{\cal
E}}_{\epsilon , \lambda} - {{\cal E}}_{\epsilon , 1-\lambda } ]. \eqn\UBCEA $$
{}From the fact
$$
\delta \Gamma_{{\bf b}{\bf c}} = - \int dt \int dx \alpha (t,x) \partial_{\mu}
<j_c^{\mu} (t,x) >, \eqn\UBCEAF $$
where $<j_c^{\mu}>$ is the ghost number current, the ghost number current
conservation law becomes
$$
\partial_{\mu} <j_c^{\mu}(t,x)> =  {\cal E}_{\epsilon ,\lambda } - {\cal
E}_{\epsilon , 1 - \lambda}  .  \eqn\GHOST $$
By using \ENERGYDC , \ENERGYDB\ and relativistic generalization
$-{\partial_1}^2 \phi \rightarrow ( {\partial_0}^2 - {\partial_1}^2 ) \phi =
-\sqrt{-g}R$, we obtain the well known anomalous conservation law
$$
\partial_{\mu} <j_c^{\mu}> = {Q \over 4 \pi} \sqrt{-g} R , \eqn\GHOSTA $$
where the background charge $Q=\epsilon (1-2\lambda)$.
Thus the non-Hermitian regularization factor leads the ghost number anomaly.
It is interesting that the ghost number anomaly arises from the total
derivative term $ {\partial_1}^2 \phi$ which has given no contribution to the
conformal anomaly.

For the scalar field and spinor field (Majorana fermions), the regularization
factors in \ENERGYS\ and \ENERGYSP\ are Hermitian, thus the vacuum energies
\ENERGYPX\ and \ENERGYPXM\ do not change under the U(1) transformation
$ f_k \rightarrow e^{i \alpha (t,x)} f_k$,
$f_k^{\ast} \rightarrow e^{-i \alpha (t,x)} f_k^{\ast}$.
Therefore the conservation law of U(1) current is anomaly free.

\endpage

\centerline{\fourteenrm ACKOWLEDGEMENTS}

The author is grateful to H. Kunitomo for valuable discussions and clarfying
remarkes.
He also would like to thank H. Itoyama for careful reading of the manuscript.

\refout
\end